\documentclass[aps,prb,reprint,showpacs,superscriptaddress,groupedaddress]{revtex4-1}
\usepackage{graphicx}
\usepackage{amsmath}
\usepackage{amssymb}
\usepackage{dcolumn}
\usepackage{dsfont}
\usepackage{latexsym}
\usepackage{rotating}
\usepackage{color}
\usepackage{latexsym}
\usepackage{bbm}
\usepackage{subfigure}
\usepackage{float}
\usepackage{epsfig}
\usepackage{psfrag}
\usepackage{natbib}
\usepackage{bm}
\usepackage{amsthm}
\usepackage{eucal}
\usepackage{mathrsfs}
\usepackage{caption}

\usepackage{color} 

\usepackage{hyperref}
\hypersetup{
colorlinks=true,final=true,
        linkcolor=red,
        citecolor=blue,
        filecolor=blue,
        urlcolor=blue}

\begin{document}

\title{A Dynamical Mean-field Study of LaNiO$_3$}

\author{Debolina Misra$^{1}$} \email{debolina@phy.iitkgp.ernet.in}
\affiliation{$^{1}$Department of Physics, Indian Institute of Technology Kharagpur, Kharagpur 721302, India.}

\begin{abstract}
While most of the rare-earth nickelates exhibit a temperature-driven sharp metal-insulator transition,LaNiO$_3$ is the only exception remaining metallic down to low temperatures. Using local density approximation as an input to dynamical mean-field calculation,metallic properties of bulk LaNiO$_{3}$ is studied. The DMFT calculations indicate that the system is a correlated Fermi liquid with an enhanced effective mass. The possibility of a pressure-driven metal-insulator transition in the system is also suggested, which can be verified experimentally.
\end{abstract}

\maketitle
\vspace{-1.0em}
\section{Introduction}
\vspace{-0.5em} 
\noindent Perovskite nickelates, with generic formula RNiO$_{3}$, where R is generally a trivalent rare-earth atom (La, Nd, Pr, Sm etc.)~\cite{Catalan2008}, form a series of fascinating compounds with unconventional electronic and magnetic properties. Except for LaNiO$_{3}$, which is a paramagnetic metal at all temperatures, the ground state of rare-earth (RE) nickelates is antiferromagnetic (AFM) insulator. Most of the RE nickelates undergo a temperature-driven first order metal-insulator transition (MIT), from a high temperature paramagnetic (PM) metal to a low temperature AFM insulator, at a particular temperature T$_{MI}$, which increases systematically with decrease in the atomic radii of the rare-earth ions~\cite{Catalan2008, Medarde1997, DDS1994}.
Amongst the nickelates, the most technologically relevant material till date, is LaNiO${_3}$. In recent years it has become a popular choice as an electrode, especially for ferroelectric thin film devices including ferroelectric capacitors and non-volatile memories~\cite{Junzhu2006, Lisun1997, Chunwang2008, Dobin2003}. Also from the structural point of view, it differs from other members of the same series: while other members have orthorhombic structure, LaNiO${_3}$ has a rhombohedral symmetry, described by R$\bar{3}$c space group. The metal-insulator transitions seen in other nickelates are structurally correlated with the crystal tolerance factor ${\it{t}}$ which is defined as the ratio between the R-O and Ni-O bond distances. The value of ${\it{t}}$ is one for an ideal cubic structure and less for the distorted ones~\cite{Catalan2008, Medarde1997}. Among all the RE nickelate compounds, LaNiO${_3}$ has the maximum tolerance factor with ${\it{t}}=$0.94~\cite{Laumillis, Hamada1993}, and hence the least distorted crystal structure. As we move from La,Pr,Nd to Eu,Y compounds, the radius of the rare-earth atom decreases, and to accommodate the size-mismatch in the unit cell, the NiO$_6$ octahedron tilts~\cite{Medarde1997}. The tilting is minimum for the first few nickelate compounds (La,Nd,Pr) and the structural distortion is claimed to have minimum effect on their electronic properties. On the other hand, for the RE ions with smaller radii, the distortion may have some role to play in the metal-insulator transitions they exhibit. Hence the claim that the MIT (or its absence, in case of LaNiO$_3$) is primarily dictated by the electronic correlation vis a vis bandwidth~\cite{Medarde1997, DDS1994}.

Theoretical models and phenomenological insights developed till date deal mostly with LaNiO$_3$/LaAlO$_3$ superlattice or heterostructures~\cite{joonhan, joonhan_2, ariadna}. Some earlier reports, based on the first-principle density functional theory calculations of the electronic stucture of NiO~\cite{anisimov_1993} and RNiO$_3$ (R$=$Na, Pr, Sm, Y, Eu, Lu and Ho) compounds~\cite{parkmillis2012, mizo_1993, mizo_2000, giova_prl}, and the physics behind their metal-insulator transitions, are available. Using density functional theory (DFT) on bulk LaNiO$_3$ and its heterostucture with LaAlO$_3$, it was shown~\cite{joonhan} that the electronic structure changes with the decrease in dimension and strain, and the orbital polarizations in these two cases are just the opposite. 

The ground state of LaNiO${_3}$ is studied here in detail using dynamical mean-field theory (DMFT). Iterated perturbation theory (IPT) approximation is used as the impurity solver for the DMFT self-consistency equation. The rest of the paper is organised as follows. In the next section, an LCAO calculation to model the existing local density approximation (LDA) density of states (DOS) used as input to the LDA+DMFT (IPT) is discussed. Using the Green's function from the LDA+DMFT calculations, the temperature dependent transport properties are calculated. In section {\bf III} the DMFT results are discussed and analysed. Section {\bf IV} is for conclusions.

\vspace{-1.0em}
\section{Methods and Formalism}
\vspace{-0.5em}
\label{Methods and Formalism}
While bulk LaNiO$_3$ has a small rhombohedral distortion, the band structure of rhombohedral LaNiO$_3$ is well explained by the zone folding of a cubic Brillouin zone into a rhombohedral Brillouin zone~\cite{re_prb} which, in turn, allows one to adopt a pseudo-cubic notation for studying the electronic structure. An LCAO band-structure calculation for the LDA density of states is performed assuming a pseudo-cubic structure, as described in previous reports~\cite{Junzhu2006, Hamada1993, re_prb}. In the cubic structure the La atom sits at the corner, Ni atom at the body-centre and the O atom at the face-centre positions. In LaNiO$_3$ the nominal configuration is Ni d$^7$ with a fully filled t$_{2g}$ band and a quarter-filled e$_g$ band~\cite{Hamada1993, hanmillis, gou2011, Dan_2010}. The nearest-neighbour interaction between Ni 3d and O 2p orbitals are taken into account. While $\sigma$ bonds are the strongest covalent bonds, resulting from a face-to-face overlap of two atomic orbitals, $\pi$ bonds are much weaker compared to $\sigma$ bonds, as they are formed due to a side-by-side overlap of atomic orbitals. Hence only $\sigma$ bondings are considered to study the LaNiO$_3$ system. The tight-binding parameters are taken from an earlier report~\cite{DDS1994}. Considering two Ni e$_g$ orbitals and three p orbitals of Oxygen atom, a 5$\times$5 LCAO Hamiltonian was constructed, which in turn gives five energy bands. This band structure result matches excellently with the one reported earlier~\cite{Hamada1993} for the relevant bands near the Fermi level (FL). The LCAO DOS for the bands closest to the Fermi level is calculated. Among those five bands, two e$_g^*$ bands which are closest to the FL and formed out of anti-bonding combination between Ni e$_g$ and O 2p bands, are found out to be degenerate at k=0 point\cite{dm_ptj}. While one band with predominantly d$_{x^2-y^2}$ character, is filled with 0.96 electrons, the other one, the nominally d$_{3z^2-r^2}$ band has a filling of 0.04 electrons\cite{dm_ptj}, which once again matches excellently with the earlier report~\cite{Hamada1993}. 

The paramagnetic LaNiO${_3}$~\cite{Stewart} system, with the half-filled e$_g^*$ band can be well-described by the half-filled Hubbard model. The single-band half-filled Hubbard model is given by
\begin{equation}
\begin{split}
&{\cal H}=-\sum_{<ij>,\sigma}t_{ij}(c_{i\sigma}^{\dagger}c_{j\sigma}+h.c.)+U\sum_in_{i\uparrow}n_{i\downarrow}
\end{split}
\end{equation}
\noindent Where t$_{ij}$ is the amplitude of hopping of electrons between nearest neighbour sites $i$ and $j$, and U is the effective on-site Coulomb repulsion, taken as a parameter with typical values appropriate for RE nickelates. The operator c$_{i\sigma}^{\dagger}$ (c$_{i\sigma}$) creates (annihilates) an electron of spin $\sigma$ at the $i$-th site.

The DMFT results were calculated, using the same half-filled $e_g^*$ band mentioned above. The DMFT approach is one of the most appropriate techniques to study the strongly correlated systems as it takes full account of local temporal fluctuations. The essential idea is to replace the lattice model by a single-site quantum impurity problem embeddded in a self-consistently determined bath~\cite{Georges1996}. It becomes exact in case of large lattice coordination number $z$. Then the hopping term $t$ is scaled to $t^*/z$ to yield a sensible limit~\cite{Georges1996, BarmNSV2010}; where $t^*$ is the effective hopping integral. The retarded Green's function for the paramagnetic phase is 
\begin{equation}
\begin{split}
&{G(\omega)}=\sum_{k}\frac{1}{\omega+i\eta-\epsilon_{k}-\epsilon_d-\Sigma(\omega)}
\end{split}
\end{equation}
\noindent Where $\eta\rightarrow0^+$ and $\Sigma(\omega)$ is the real frequency self-energy, which is local within the DMFT approach~\cite{Metvoldt1989, Georges1996}.
The local retarded Green's function may also be written as
\begin{equation}
\begin{split}
&{G(\omega)}=H[\gamma(\omega)]
\end{split}
\end{equation}
\noindent Where $\gamma=\omega+i\eta-\epsilon_d-\Sigma(\omega)$ and $H(z)$ is the Hilbert transform of $z$, given by
\begin{equation}
\begin{split}
&{\cal H}(z)=\int{d\epsilon}\frac{\rho_0(\epsilon)}{z-\epsilon}
\end{split}
\end{equation} 
\noindent with $\rho_0$ as the non-interacting density of states.
The self-consistency condition in DMFT demands that the lattice self energy is same as the impurity self-energy. So the Green's function of the external bath ${\cal G}(\omega)$ can be obtained from Dyson's equation

\begin{equation}
{{\cal G}^{-1}(\omega)}=G^{-1}(\omega)+\Sigma(\omega)
\end{equation}

\noindent The newly evolved dynamical mean field would then yield a new self-energy, and hence a new G($\omega$)~\cite{BarmNSV2010}.
The self-consistency in DMFT ensures that the local component of the Green's function coincides with the one calculated from the effective single-site action S$_{eff}$, related to the bare Green's function~\cite{Georges1996}.

The major simplification in the DMFT approach is that the vertex correction is not needed to calculate conductivity and only the elementary particle-hole bubble survives~\cite{Georges1996}. The optical conductivity is obtained from the Kubo formula and is given by,

\begin{equation}
\begin{split}
&{\sigma(\omega)}={\frac{\sigma_0}{2\pi^2}}Re\int_{-\infty}^{+\infty}{d\omega'}{\frac{n_F(\omega')-n_F(\omega+\omega')}{\omega}}\\
&\times[{\frac{G^*(\omega')-G(\omega+\omega')}{\gamma(\omega+\omega')-\gamma^*(\omega')}}-{\frac{G(\omega')-G(\omega+\omega')}{\gamma(\omega+\omega')-\gamma(\omega')}}]
\end{split}
\end{equation}
\noindent Where $\sigma_0=\frac{4\pi e^2 t^2 a^2 n}{\hbar}$; a is the lattice constant~\cite{BarmNSV2010}. The D.C conductivity can be obtained simply by using the limit $\omega\rightarrow0$. Here IPT approximation is used for solving the self-consistent impurity problem. IPT is one of the most simple yet accurate techniques to study the Hubbard model. In IPT the second order term of the perturbative expansion in U is taken into account and is given by

\begin{equation}
{\Sigma_2}(\omega)=\lim_{i\omega\rightarrow\omega^+}\frac{U^2}{\beta^2}\sum_{m,n}{\cal G}(i\omega+i\nu_m){\cal G}(i\omega_n+i\nu_m){\cal G}(i\omega_n)
\end{equation}

\noindent Where $i\nu$ and $i\omega$ are the even and odd Matsubara frequencies respectively, and the spectral function~\cite{BarmNSV2010} is
\begin{equation}
{\cal G}(i\omega_n)=\int_{-\infty}^{+\infty}d\omega'\frac{D(\omega')}{i\omega_n-\omega'}
\end{equation}
\noindent Where $D(\omega')=-Im{\cal G}/\pi$. 

\noindent Then the Matsubara summation is carried out with the analytical continuation $i\omega_n\rightarrow\omega+i\eta$. After calculating the imaginary part of self-energy, the real part is found out by Kramers-Kroning transformation. The angle resolved photo emission spectra (ARPES) for LaNiO$_3$, are calculated along three different symmetry directions, namely $\Gamma-$X, X-M and M-R directions of the cubic Brilloiun zone, using the DMFT spectral function. ARPES is one of the most powerful methods to study the electronic structure of a solid. For strongly correlated systems it is very effective in elucidating the connection between electronic and magnetic properties of a solid and effects of correlation thereon. ARPES intensity is basically the convolution of A($k,\omega$)$\times$f($\omega$) with IB($\omega$) where f($\omega$) is the Fermi function and IB($\omega$) is the instrumental broadening. Instrumental resolution for this system is taken from earlier reports~\cite{valla, Damascelli}. A($k,\omega$) is the quasi-particle spectral function given by~\cite{valla, Damascelli}:

\begin{equation}
\begin{split}
&{A(k,\omega)}=-\frac{Im\Sigma(\omega)}{\pi[\omega-\epsilon_k-Re\Sigma(\omega)]^2+[Im\Sigma(\omega)]^2}
\end{split}
\end{equation}

\noindent Where $\Sigma(\omega)$ denotes the self-energy and $\epsilon_k$ is the LDA energy spectrum.

\vspace{-1.0em}
\section{Results and Analysis}
\begin{figure}[h]
\centering
\psfig{file=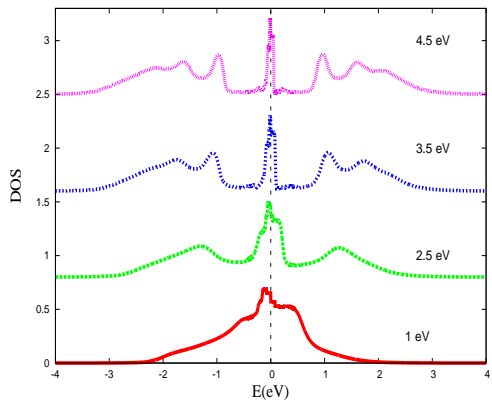, height=0.37\textwidth, width=0.3\textwidth, angle =-90}
\caption{(Color online) Evolution of DMFT DOS for different U values.}
\label{lno_dmftdos}
\end{figure}

\vspace{-0.5em}
\label{discussion}
 The bandwidth of the almost half-filled e$_g^*$ band, crossing the Fermi level, is found out to be 2.64 eV\cite{dm_ptj}. As the value of U$_{dd}$ for LaNiO${_3}$ is 4.7$\pm$0.5eV~\cite{DDS1994}, so $U\ge W$ puts the system into the strongly correlated category. This ${x^2-y^2}$ band is used for DMFT calculations.  Implementing the above mentioned technique, the spectral functions for different interaction strengths, are calculated numerically. The evolution of DMFT DOS for different U values is shown in Fig.~\ref{lno_dmftdos}. Even for a very large interaction strength, there is still a non-vanishing DOS at the Fermi level, which clearly indicates the metallic state of the system.

\vspace{-1.0em}
\subsection{Self-energy and the enhancement of effective mass}
\begin{figure}[h]
\centering
\psfig{file=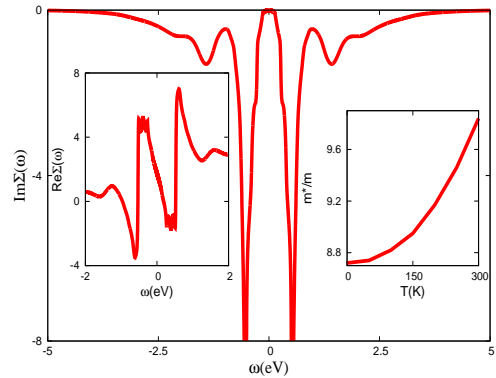, height=0.37\textwidth, width=0.28\textwidth, angle=-90}
\caption{(Color online) Imaginary part of self-energy 
Im${\Sigma(\omega)}$ calculated via DMFT for U=4 eV. The left inset shows the real part of self-energy and the right inset shows the variation of effective mass with temperature.}
\label{self_energy}
\end{figure}

\vspace{-1.0em}
Both the real and imaginary parts of the self-energy $\Sigma(\omega)$ are calculated for different values of Coulomb repulsion. The imaginary part of the self-energy for a reasonably strong $U$ value is shown in Fig.~\ref{self_energy}. Variation of the real part of self-energy Re${\Sigma}(\omega)$ with $\omega$ is shown in the left inset of Fig.~\ref{self_energy}. Im${\Sigma}$ shows a quadratic variation with energy close to the Fermi level and no pole is observed at $\omega=$0. The quadratic variation of  Im${\Sigma}(\omega)$ with $\omega$ and the absence of a pole at $\omega=$0 clearly indicate the metallic nature, albeit strongly correlated, of the system under consideration. The $\omega^2$ dependence of the imaginary part of the self-energy is consistent with the results of earlier reports~\cite{parkmillis2012,re_prb}.

The effective mass of the electron is easily calculated within DMFT as it is related to the quasi-particle residue $Z$ of the Green's function via the equation 
\begin{equation}
\begin{split}
&{Z}=\frac{1}{1-\frac{\partial Re\Sigma(\omega)}{\partial \omega}}
=\frac{m_0}{m^*}
\end{split}
\end{equation}

\noindent at $\omega=$0, where m$_0$ is the free electron mass. The effective mass was calculated for various U values. At U=3.5 eV the effective mass came out to be 8.71m$_0$ which is close to the effective mass reported earlier~\cite{re_prb,s_shin,sakai_2002} corroborating the range of typical U-values relevant for this system. Considerable enhancement of the effective mass also indicates the correlated nature of the metallic state. The temperature dependence of effective mass was also studied over a wide range of temperature. However no significant variation in effective mass has been observed with change in temperature for the range studied, indicating that the metallic nature of the system does not change with temperature. The right inset of Fig.~\ref{self_energy} shows the narrow range of variation of the effective mass of the electrons with temperature for U$=$3.5 eV.

\vspace{-1.0em}
\subsection{Photoemission spectra}
\begin{figure}[h!]
\begin{subfigure}[]
\centering
\psfig{file=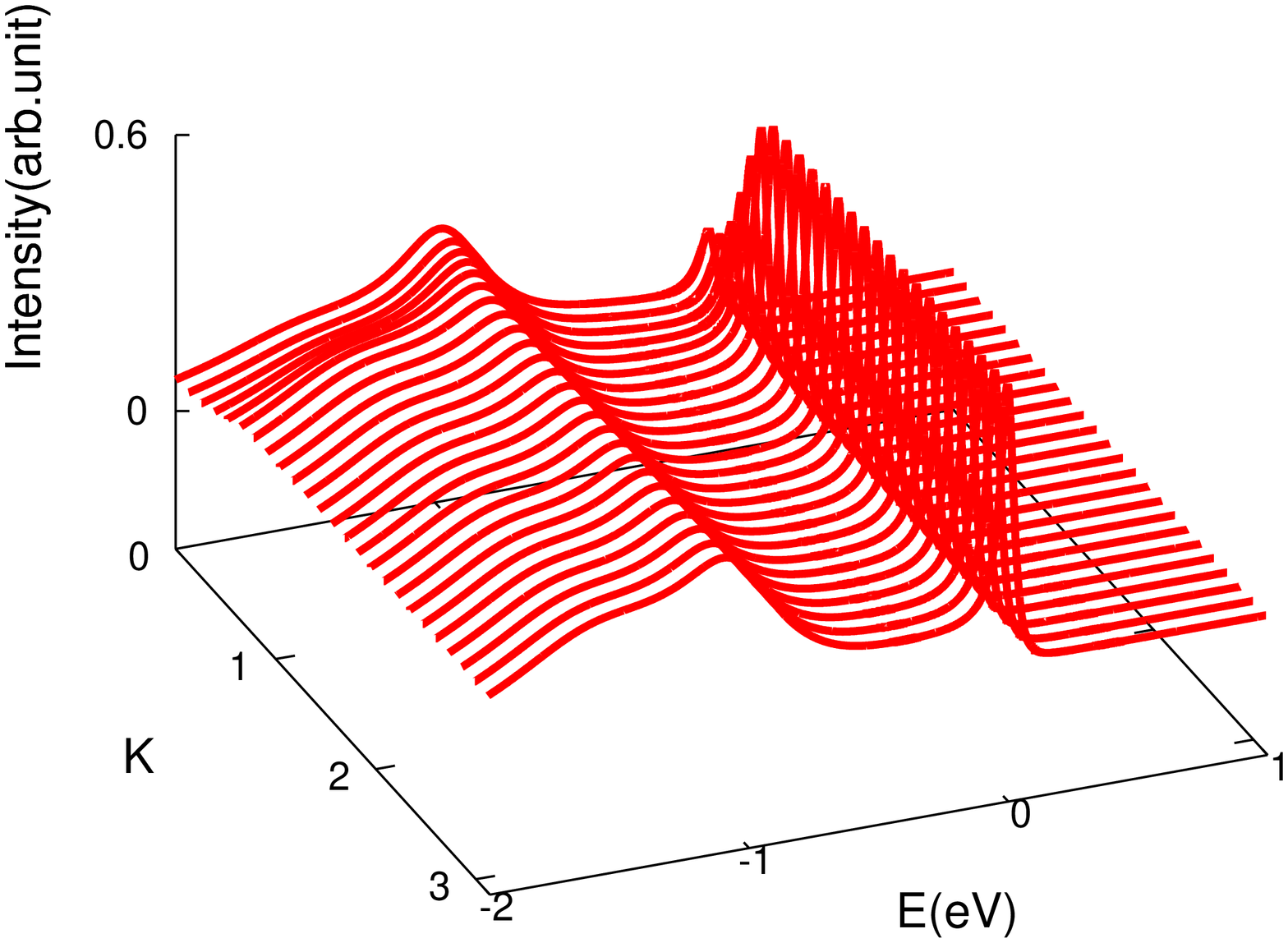, height=0.37\textwidth, width=0.3\textwidth, angle=-90}
\end{subfigure}

\begin{subfigure}[]
\centering
\psfig{file=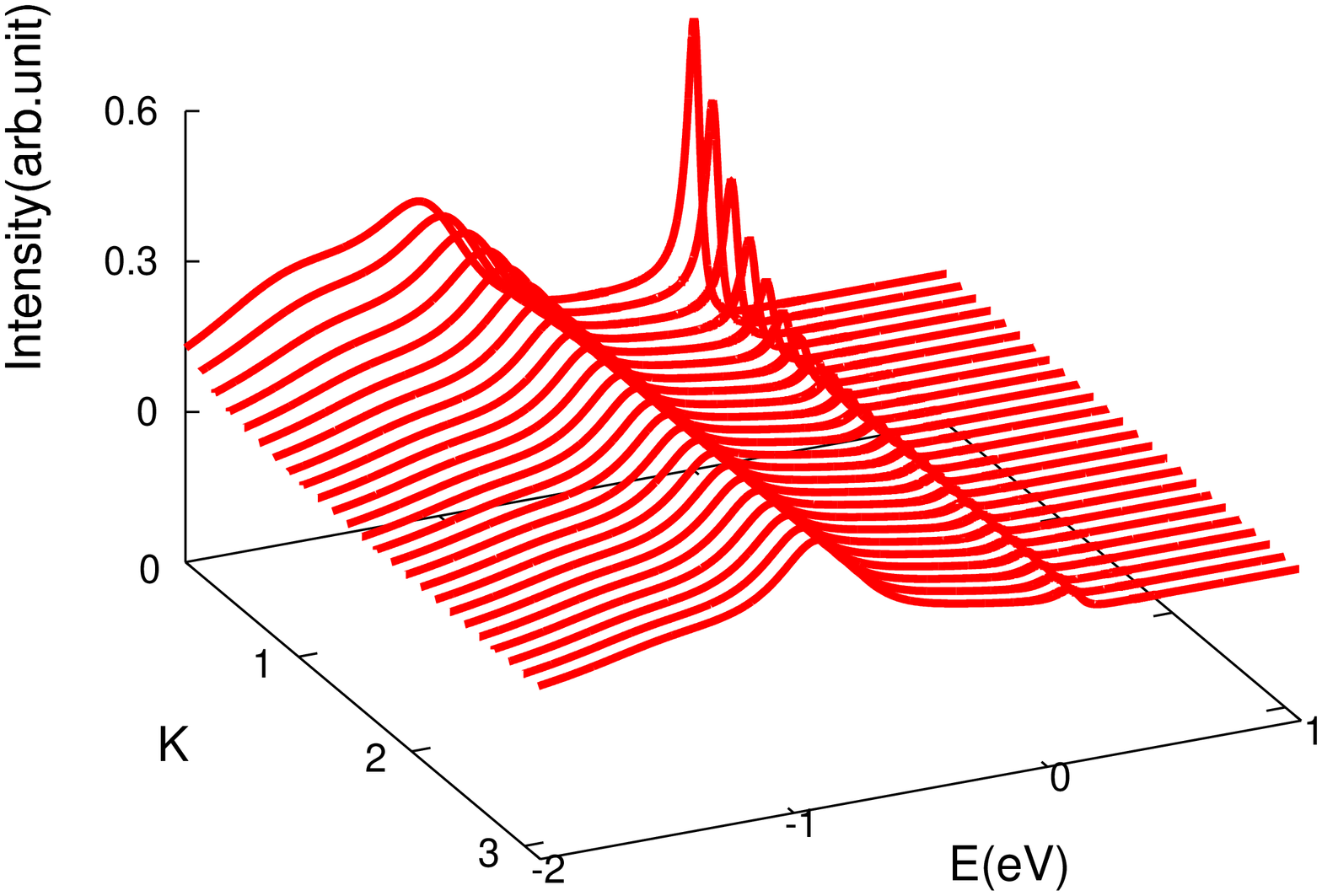, height=0.37\textwidth, width=0.3\textwidth,angle=-90}
\end{subfigure}

\begin{subfigure}[]
\centering
\psfig{file=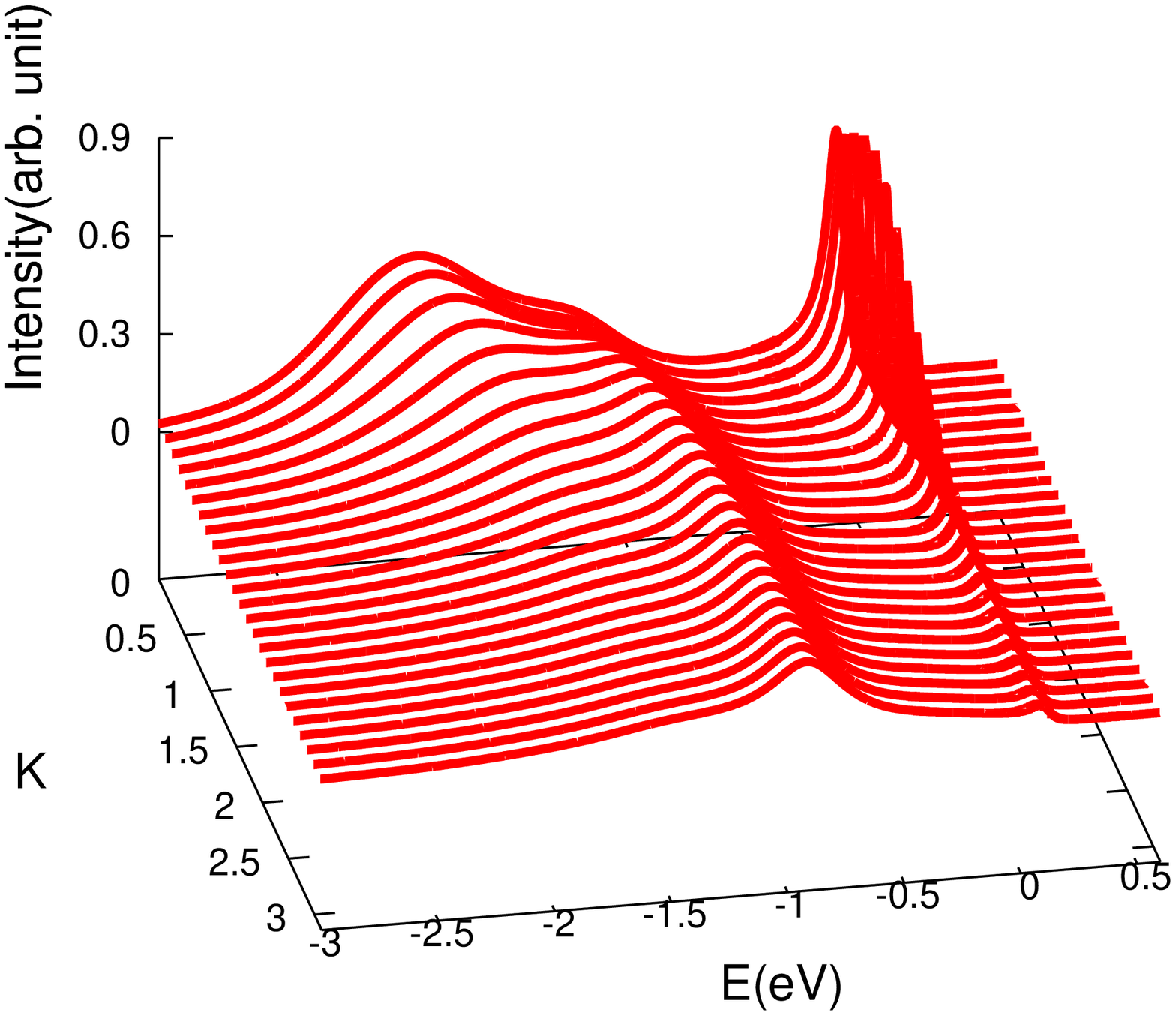, height=0.37\textwidth, width=0.3\textwidth,angle=-90}
\end{subfigure}
\caption{(Color online) Theoretically computed ARPES along (a)$\Gamma$-X(b)X-M and(c)M-R directions for U=3.5 eV.}
\label{pes}
\end{figure}

\vspace{-1.0em}
The energy distribution curves (EDC) along three major symmetry directions ($\Gamma$-X, X-M and M-R) are shown in Fig.~\ref{pes}. In Fig.~\ref{pes}a, the large peak appearing near the $\Gamma$ point is from the electron pocket around it.  The dispersion curve of cubic LaNiO$_3$ has an electron pocket centred around the $\Gamma$ point~\cite{re_prb}. This electron pocket seems to be responsible for the huge peak proximate to k$=$0~(Fig.\ref{pes}a). Another peak appearing at around -1.2 eV is attributed to the inter Hubbard subband transition: the band edges of the lower and upper Hubbard subbands formed at $\pm$0.6 eV as shown in Fig.~\ref{lno_dmftdos}.

\noindent Fig.~\ref{pes}b shows the ARPES within DMFT along X-M symmetry direction. The peak appearing close to 0.6 eV in Fig.~\ref{pes}b is due to the excitation between the Hubbard subband and the Fermi level. Due to the almost non-dispersive nature of the energy dispersion curve~\cite{re_prb} in this direction, variation in intensity along ${\bf k}$ direction is negligible, which is reported in earlier experiments~\cite{re_prb}. In Fig.~\ref{pes}c, a rise in the intensity has  been observed near -0.6 eV and owes its origin to the existence of the lower Hubbard subband at around -0.6 eV. Another increase in intensity around the $\Gamma$ point is also seen at the energy 0.25 eV. This intensity rise at 0.25 eV mathces excellently with an earlier report~\cite{re_prb}, where it was claimed that this feature is a typical characteristics of a strongly correlated system.

\vspace{-1.0em}
\subsection{Specific heat}
\vspace{-1.0em}

\begin{figure}[h]
\centering
\psfig{file=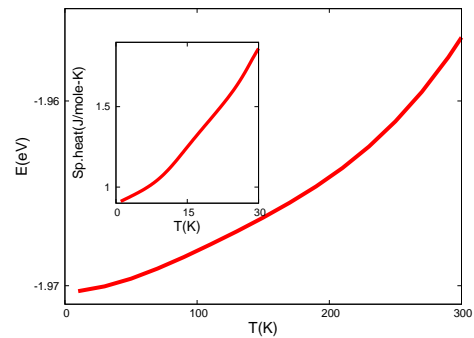, height=0.35\textwidth, width=0.25\textwidth, angle=-90}
\caption{(Color online) Variation of internal energy with temperature. The inset shows the variation of low-temperature specific heat of the system with temperature, for U$=$3.5 eV.}
\label{int_sp}\vspace{-0.5em}
\end{figure}

\noindent Internal energy of the system has been computed for different values of U. Variation of internal energy with temperature for U=3.5 eV is shown in Fig.~\ref{int_sp}. The internal energy of the system does not change much with temperature, as shown in Fig.~\ref{int_sp}. For the whole temperature range observed, the curvature of the curve remains positive, indicating that the specific heat of the system is growing with temperature as expected in a metal.

The inset of Fig.~\ref{int_sp} shows the low-temperature specific heat of the system, calculated at U=3.5 eV. The DMFT specific heat data match very well with earlier reports~\cite{sanch}; the specific heat varies linearly with temperature upto T$=$10K. The variation of specific heat with temperature fits well with the usual equation $C(T)=\gamma T +\beta T^3$ as mentioned earlier~\cite{sanch}. The specific heat coefficient $\gamma$ turns out to be 14.57 mJ/mol-K$^2$, which matches again with the $\gamma$ values reported for bulk LaNiO$_3$(13.04 mJ/mol K$^2$ ~\cite{sanch} and 15 mJ/mol K$^2$ ~\cite{raj}) earlier. 

\vspace{-1.0em}
\subsection{Transport}
\vspace{-1.0em}

\noindent
\vspace{-1.0em}
\subsubsection{Optical conductivity}
\vspace{-1.0em} 

\begin{figure}[h]
\centering
\psfig{file=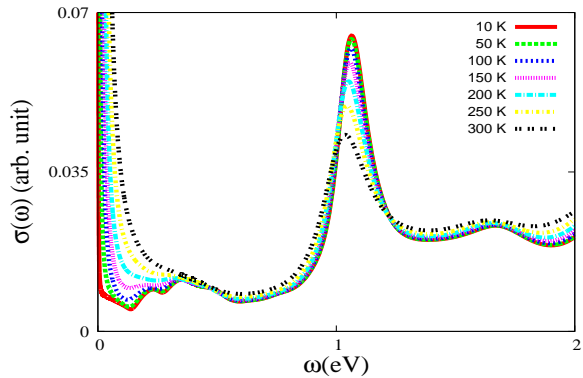, height=0.45\textwidth, width=0.28\textwidth, angle =-90}
\caption{(Color online)Variation of real part of optical conductivity with temperature.}
\label{opt_T}
\end{figure}

As mentioned above, the optical conductivity has been calculated using Kubo formula for different $U$ and at different temperatures. Variation of optical conductivity with temperature for U=3.5 eV, is shown in Fig.~\ref{opt_T}. For all temperatures a Drude peak at very low energies, typical characteristic of metals, is evident. As temperature increases, a spectral weight transfer from higher to lower energy has been observed. Additionally there is a broad hump-like feature at around 1.2 eV. From Fig.~\ref{lno_dmftdos} it is clear that, the formation of two Hubbard subbands starts as the interaction strength is raised above 2 eV. At U$=$3.5 eV the Hubbard subbands can be clearly seen with edges around $\pm$0.6 eV. Hence the shoulder like feature in the optical conductivity spectra, appearing at 1.2 eV, is attributed to the excitation across the Hubbard subbands. An earlier report on the optical conductivity of Hubbard model~\cite{kontani_2006} claimed that the hump like feature is the signature of metallic phase and was named a pseudo-Drude peak.  

\begin{figure}[h!]
\begin{subfigure}[]
\centering
\psfig{file=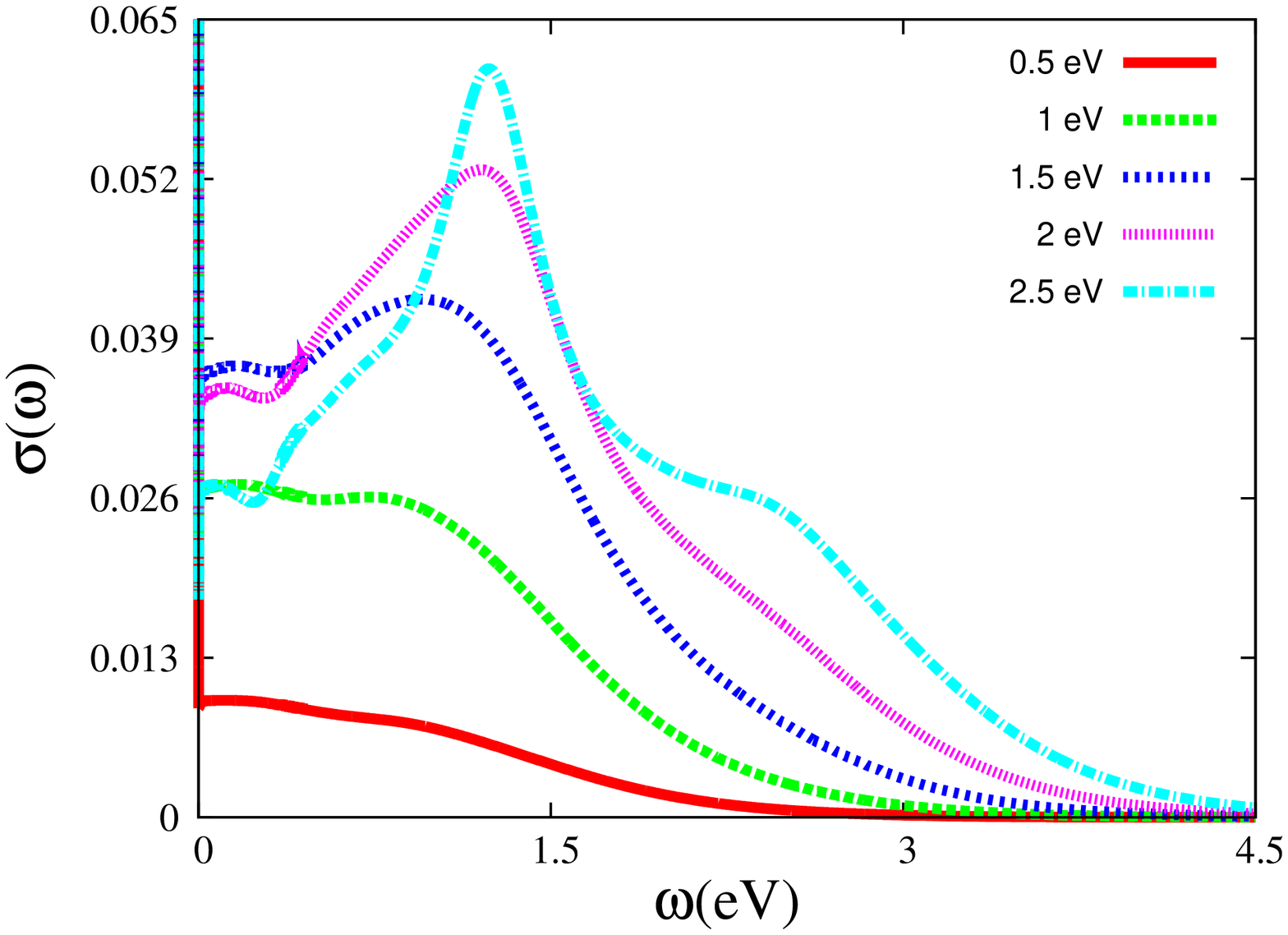, height=0.35\textwidth, width=0.28\textwidth,angle=-90}
\end{subfigure}

\begin{subfigure}[]
\centering
\psfig{file=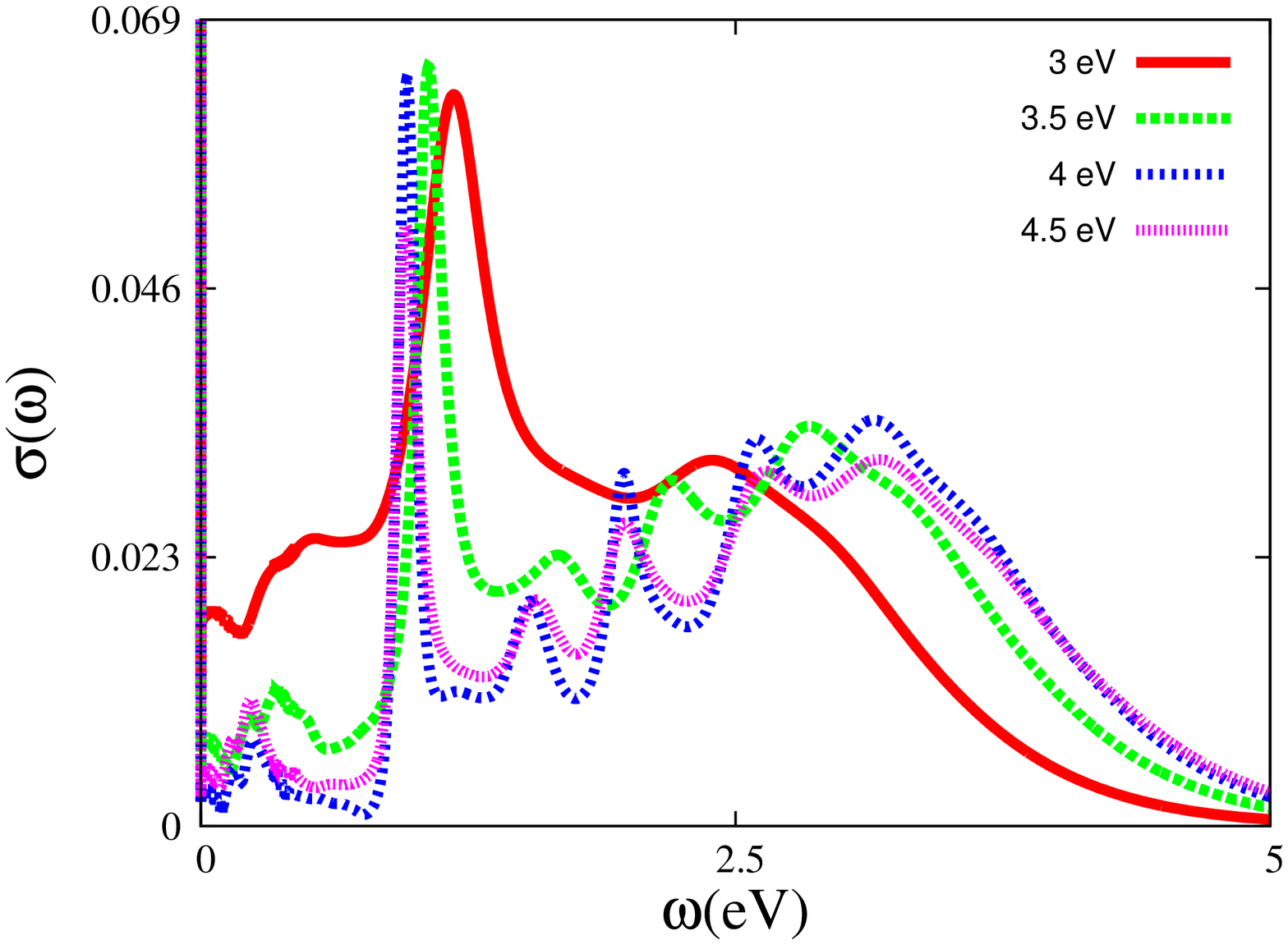, height=0.35\textwidth, width=0.28\textwidth,angle=-90}
\end{subfigure}
\caption{(Color online) Real part of optical conductivity obtained at T$=$0K for (a)low and (b)high U values respectively.}
\label{opt_u}
\end{figure}

The claim of optical excitation between the Hubbard subbands can be substantiated from the optical conductivity spectra obtained at T=0 K, for different U values. The variation of $\sigma(\omega)$ is studied for U values ranging from 0.5 eV to 4.5 eV. Fig.~\ref{opt_u}a shows the variation of optical conductivity as Coulomb repulsion U increases from a small value of 0.5 eV to a moderately high value of 2.5 eV. In Fig.~\ref{opt_u}b the change in $\sigma(\omega)$ for higher values of U is shown. In all the cases the Drude peak is present (visible only as a very sharp rise almost within the energy resolution of $\omega=0$).

\vspace{-1.0em}
\subsubsection{Resistivity}
\vspace{-1.0em} 

\begin{figure}[h!]
\centering
\psfig{file=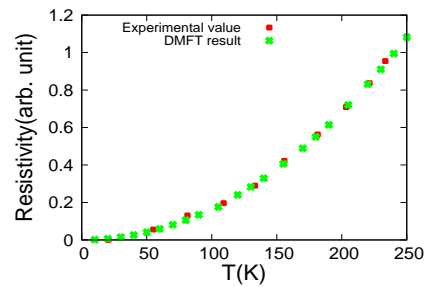, height=0.32\textwidth, width=0.21\textwidth, angle =-90}
\caption{(Color online)Comparison of resistivity obtained from DMFT, with experimental results.}
\label{res}
\end{figure}
\begin{figure}[h!]
\centering
\psfig{file=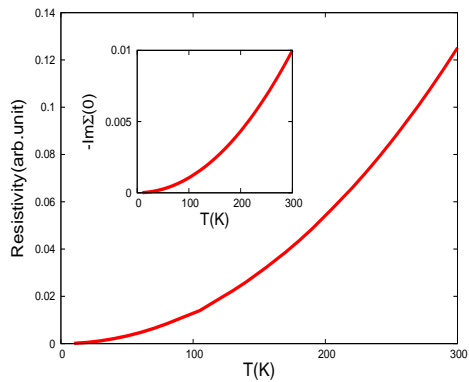, height=0.35\textwidth, width=0.28\textwidth, angle =-90}
\caption{(Color online) Variation of resistivity with temperature obtained at U=2 eV. The inset shows variation of scattering rate with temperature for the same U value.}
\label{res_scat}
\end{figure}

The temperature dependence of resistivity is shown in Fig.~\ref{res}. Resistivity shows a quadratic dependence on temperature, which is a signature of a correlated metal. Resistivity obtained within the DMFT for bulk LaNiO$_3$ matches excellently(Fig.~\ref{res}) with earlier reports~\cite{re_prb,wang_2011,s_shin}. The T$^2$ dependence of resistivity is a characteristic of electron-electron interaction~\cite{json_2010,sre_1992} in the system. The scattering rate of electrons is also calculated from the DMFT self-energies. Im$\Sigma(\omega)$ at $\omega=$0 is a measure of the scattering rate due to correlation. The variation of computed scattering rate with temperature is shown in Fig.~\ref{res_scat}. The scattering rate also shows a T$^2$ dependence, closely following experimental observations.

\vspace{-1.0em}
\subsection{Effect of Pressure}
\vspace{-1.0em}

\begin{figure}[h]
\centering
\psfig{file=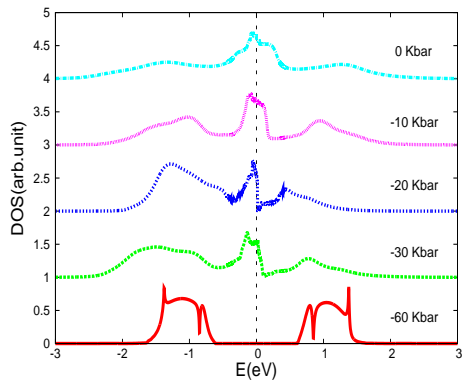, height=0.35\textwidth, width=0.28\textwidth, angle =-90}
\caption{(Color online) Effect of pressure on DMFT DOS for U=2 eV.}
\label{pres}\vspace{-0.5em}
\end{figure}

RNiO$_3$ compounds are claimed to be extremely sensitive to pressure and strain~\cite{Can_1993,obrad,ishi}. Ultra-thin films of LaNiO$_3$ show MIT due to epitaxial strain~\cite{json_2010,chakha_prl}. Effect of hydrostatic pressure on bulk LaNiO$_3$ is considered following the standard approach~\cite{harrison}, wherin pressure causes a change in bandwidth via increased overlap of wave functions and alters the hopping integrals within the lattice. As pressure is applied to a system, it changes the bandwidth following the relation ${D(P)}=D_0\exp[\gamma k_lP]$, where D(P) is the bandwidth under pressure P, D$_0$ is the same without pressure, and k$_l$ is the compressibility. The altered bandwidth changes the hopping parameters, following ${t}=t_0{\frac{D(P)}{D_0}}^2$, where t$_0$ and t are the hopping amplitudes of electrons between the nearest neighboring sites before and after the application of pressure respectively. Negative pressure can be incorporated easily as well. In reality, such negative pressure is applied through substitution by atoms of larger radii. For each value of pressure corresponding bandwidths were calculated, which in turn changed the hopping parameters and hence the band structure that goes in to DMFT input. Since LaNiO$_3$ is always metallic, the effect of negative pressure to search for MIT is studied and is shown in Fig.~\ref{pres}. It is clear from Fig.~\ref{pres} that the system is highly sensitive to (negative) pressures of even a few kilo-bars. At P=-60 Kbar, there is a gap opening at the Fermi level, leading to an insulating state for a small interaction strength U=2 eV.  As the lattice parameter increases, the orbital overlap decreases and after a critical value of pressure the system becomes insulating. This should be clearly observable in experiments, provided a suitable clean method (via, e.g., chemical substitution) of applying negative pressure can be devised.

\vspace{-1.0em}
\section{Conclusions} 
\vspace{-1.0em}
\label{conclusion}

In summary, the transport calculations of the strongly correlated metallic LaNiO$_3$ system is done in detail, using a single-orbital DMFT approach. The anti-bonding e$_g^*$ band formed by the overlap between Ni-3d and O-2p orbitals seems to be responsible for the conduction mechanism within the solid. The DMFT DOS at the Fermi level remains finite even for a large Coulomb interaction. The non-vanishing DOS at the Fermi level, even at reasonably large correlation, explains the metallic nature of the system. The resistivity data clearly show a T$^2$ dependence on temperature which indicates that LaNiO$_3$ is a Fermi liquid as also reported earlier~\cite{re_prb}, with a correlation driven enhanced effective mass of electron. Variation in optical conductivity with both temperature and interaction strength U, and the ARPES data reveal the signatures of optical excitations between Hubbard subbands and the Fermi level. A pressure-driven metal-insulator transition in the system with the application of a few kilo-bar of negative pressure is predicted. We have also performed a DMFT calculation on NdNiO$_3$~\cite{dm_to_be} and found a metal-insulator transition at around U=3.2 eV. This remarkable difference of fate of the two almost similar compounds appears to be driven by the competition of correlation and bandwidth played out subtly and owes its origin to the dynamical correlations beyond a static mean-field theory.

\vspace{-1.0em}
\section*{Acknowledgements}
\vspace{-0.5em} 
The author profusely thanks Prof. N. S. Vidhyadhiraja for his help in the DMFT work. In fact this work owes a lot on Prof. Vidhyadhiraja's DMFT help. Special thanks to Prof. A. Taraphder for valuable discussions.


\begin{thebibliography}{10}

\bibitem{Catalan2008}
G.~Catalan,
\newblock Phase Transitions {\bf 81}, 729 (2008).

\bibitem{Medarde1997}
M.~Medarde,
\newblock  J. Phys.: Condens. Matter {\bf 9}, 1679 (1997).

\bibitem{DDS1994}
D. D.~Sarma, N.~Shanthi, and P.~Mahadevan
\newblock J. Phys.: Condens. Matter {\bf 6}, 10467 (1994).

\bibitem{Junzhu2006}
Jun Zhu {\em et~al.},
\newblock Materials Chemistry and Physics {\bf 100}, 451 (2006).

\bibitem{Lisun1997}
Li Sun {\em et~al.},
\newblock J. Mater. Res {\bf 12}, 931 (1997).

\bibitem{Chunwang2008}
Chun Wang and Mark H Kryder,
\newblock Phys. Scr. {\bf 78}, 035601 (2008).

\bibitem{Dobin2003}
A.~Yu. Dobin {\em et~al.},
\newblock Phys. Rev. B, {\bf 68}, 113408 (2003).

\bibitem{Laumillis}
 B.~Lau and A.~J.~Millis,
\newblock ArXiv e-prints (2012), 1210.6693v1.

\bibitem{Hamada1993}
N.~Hamada,
\newblock J. Phys. Chem. Solids {\bf 54}, 1157 (1993).

\bibitem{joonhan}
M.~Joon Han and M.~van Veenendaal,
\newblock Phys. Rev. B {\bf 84}, 125137 (2011).

\bibitem{joonhan_2}
M.~Joon Han and M.~van Veenendaal,
\newblock Phys. Rev. B {\bf 85}, 195102 (2012).

\bibitem{ariadna}
A.~Blanca-Romero and R.~Pentcheva,
\newblock Phys. Rev. B {\bf 84}, 195450 (2011).

\bibitem{anisimov_1993}
V.~I.~Anisimov, {\em et~al.},
\newblock Phys. Rev. B {\bf 48}, 16929 (1993).

\bibitem{parkmillis2012}
H.~Park, A.~J.~Millis, and C.~A.~Marianetti,
\newblock ArXiv e-prints (2012),1206.2822v1.

\bibitem{mizo_1993}
T.~Mizokawa, {\em et~al.},
\newblock  Phys. Rev. B {\bf 52}, 13865 (1995).

\bibitem{mizo_2000}
T.~Mizokawa, D.~I.~Khomskii, and G.~A.~Sawatzky,
\newblock  Phys. Rev. B {\bf 61}, 11263 (2000).

\bibitem{giova_prl}
G.~Giovannetti, {\em et~al.},
\newblock  Phys. Rev. Lett. {\bf 103}, 156401 (2009).

\bibitem{re_prb}
R.~Eguchi, {\em et~al.},
\newblock ArXiv e-prints (2009), 0903.1487v1.

\bibitem{hanmillis}
M.~J.~Han, X.~Wang, C.~A.~Marianetti, and A.~J.~Millis,
\newblock Phys. Rev. Lett., {\bf 107}, 206804 (2011).

\bibitem{gou2011}
G.~Gou, I.~Grinberg, A.~M.~Rappe, and J.~M.~Rondinelli,
\newblock Phys. Rev. B,{\bf 84}, 144101 (2011).

\bibitem{Dan_2010}
D.~G.~Ouellette, {\em et al.},
\newblock Phys. Rev. B, {\bf 82}, 165112 (2010).

\bibitem{dm_ptj}
Debolina~Misra,
\newblock Phase Transitions (2014), DOI:10.1080/01411594.2013.853767.

\bibitem{Stewart}
M.~K.~Stewart, {\em et~al.},
\newblock Phys. Rev. B, {\bf 83}, 075125 (2011).

\bibitem{Georges1996}
A.~Georges,{\em et~al.},
\newblock Rev. Mod. Phys., {\bf 68}, 13 (1996).

\bibitem{BarmNSV2010}
H.~Barman and N.~S.~Vidhyadhiraja,
\newblock Int. J. Mod. Phys. B, {\bf 25}, 2461 (2011).

\bibitem{Metvoldt1989}
W.~Metznerand and D.~Vollhardt,
\newblock Phys. Rev. Lett. {\bf 62}, 324 (1989).

\bibitem{valla}
T.~Valla,{\em et~al.},
\newblock Phys. Rev. Lett. {\bf 83}, 2085 (1999).

\bibitem{Damascelli}
Andrea Damascelli,
\newblock Physica Scripta, {\bf T109}, 61 (2004).

\bibitem{s_shin}
S.~Shin,
\newblock Materials Science: Electronic \& Magnetic Properties, {\bf Spring8}. 

\bibitem{sakai_2002}
A.~Sakai, G.~Jheng, and Y.~Kitaoka,
\newblock J. Phys. Soc. Jpn., {\bf 71}, 166 (2002).

\bibitem{sanch}
R.~D.~Sanchez, {\em et al.},
\newblock Journal of Alloys and Compounds, {\bf 191}, 287 (1993).

\bibitem{raj}
K.~P.~Rajeev, G.~V.~Shivashankar, and A.~K.~Raychaudhuri,
\newblock Solid State Commun. {\bf 79} 591 (1991).

\bibitem{kontani_2006}
T.~Mutou and H.~Kontani,
\newblock Phys. Rev. B {\bf 74}, 115107 (2006).

\bibitem{wang_2011}
Y.~Wang, {\em et~al.},
\newblock Bull. Mater. Sci., {\bf 34}, 1379 (2011).

\bibitem{json_2010}
J.~Son, {\em et~al.},
\newblock Appl. Phys. Lett, {\bf 96}, 062114 (2010).

\bibitem{sre_1992}
K.~Sreedhar, {\em et~al.},
\newblock Phys. Rev. B, {\bf 46}, 6382 (1992).

\bibitem{Can_1993}
P.~C.~Canfield, {\em et al.},
\newblock Phys. Rev. B, {\bf 47}, 12357 (1993).

\bibitem{obrad}
X.~Obradors, {\em et al.},
\newblock Phys. Rev. B, {\bf 47}, 12353 (1993).

\bibitem{ishi}
S.~Ishiwata, {\em et al.},
\newblock Phys. Rev. B, {\bf 72}, 12357 (2005).

\bibitem{chakha_prl}
J.~Chakhalian, {\em et al.},
\newblock Phys. Rev. Lett, {\bf 107}, 116805 (2011).

\bibitem{harrison}
Walter A.~ Harrison,
\newblock Solid State Theory, Dover (1980).

\bibitem{dm_to_be}
D.~Misra, N.~S.~Vidhyadhiraja, and A.~Taraphder,
\newblock To be submitted.


\end{thebibliography}
\end{document}